\documentclass[letterpaper, aps, showpacs,twocolumn]{revtex4}
\usepackage{amsmath}
\usepackage{amssymb}
\usepackage{latexsym}
\usepackage[dvips]{graphics}
\usepackage{hyperref}

\begin{document}

\title{Generalized grand-canonical ensemble theory for interacting Bose-Einstein systems}
\author{Alexander Jurisch and Jan-Michael Rost}
\affiliation{Max-Planck-Institut f\"ur Physik komplexer Systeme\\
N\"othnitzerstr. \!\!38, 01187 Dresden, Germany}
\begin{abstract}
We use the maximum information principle to include particle-interaction into the grand-canonical theory of BECs. The inclusion of the particle-interaction elucidates why thermodynamic calculations for BECs by the grand-canonical ensemble for the non-interacting case are in coincidence with up to date experimental results. However, in our generalized theory we can show that a BEC experiences a real and abrupt phase-transition in contrast to the smooth phase-transition predicted by the non-interacting grand-canonical ensemble. In addition, we discuss possible effects due to a thermal environment and the experimental probing from a general point of view.
\end{abstract}
\maketitle
\section{Introduction}
The prediction of a macroscopical, condensed state for bosonic systems in \cite{Ein, Bos} started an extremely fruitfull theoretical development in order to describe this phenomenon in all details. The theory of Bose-Einstein condensation has been worked out macroscopically within the framework of the canonical and grand-canonical ensemble, e.g. \cite{Ein, Bos, deG, Zif, Bag}, but only for cases without interaction between particles. Particle-interaction due to s-wave scattering, on the other hand, was included in microscopical approaches related to quantum field theory at zero temperature. Amongst the microscopical approaches, the standard references are the theory by Bogoliubov \cite{Bog} for the quasi-particle picture and the work of Gross \cite{Gro} and Pitaevskii \cite{Pit} for the inclusion of particle-interaction into the framework of the Schr\"odinger theory. Contact between the interacting field-theoretical and the non-interacting ensemble approaches is usually made by including microscopically calculated spectral corrections into the statistical description of the many-particle system by various techniques that are reviewed in, e.g., \cite{GriSnoStr, Cornell}.

Our work will pursue a completely different, exclusively macroscopical ansatz for an equilibrium ensemble, that also may be called a holistic one. Our ansatz is based on the well defined fact that all available and necessary information about a macroscopical physical system can be obtained from its entropy. The meaning of entropy as a measure for information was first recognized by Shannon \cite{Sha} and was brought back to its physical context by Jaynes \cite{Jay}. In \cite{Jay} a variational principle was developed that allows us to derive the relevant physical distribution functions under several constrains imposed on the entropy by the method of Lagrange-parameters. This variational principle is also known as the maximum information principle (MIP), e.g. \cite{Haken}. By the MIP, all statistical ensembles known are derived most easily when the appropriate constrains are applied.

In our approach we will use the MIP to include the particle-interaction for the BEC-case into the statistical grand-canonical ensemble description in a non-perturbative way. The generality of the MIP ensures, that only the most elementary knowledge about the BEC-system is required: the Hamiltonian, the particle-conservation and the symmetry of the many-particle wave-function.

We further will discuss environmental effects onto the generalized BEC-ensemble, that are directly related to its non-linear structure. We will treat the environment in the sense of equilibrium statistics which is valid for a system after cooling-down and relaxation to equilibrium have come to an end and only quasi-stationary interactions between system and environment are left.

\section{Bose-Einstein-statistics including particle-interaction}
The quantum system describing interacting bosons in terms of the density-matrix is given by the Hamiltonian
\begin{equation}
\mathcal{H}\,=\,{\rm{tr}}[\hat{H}]\,=\,{\rm{tr}}[\hat{H}_{0}\,\hat{\rho}\,+\,g\,\hat{\rho}^2]\quad,
\label{hamiltonian}\end{equation}
where $\hat{H}_{0}\,=\,\hat{T}\,+\,\hat{U}$ is the non-interacting part of the system containing the kinetic energy $\hat{T}$ and the external potential $\hat{U}$.

We abbreviate the entropy $S$ by $S\,=\,{\rm{tr}}[\hat{S}]\,=\,-{\rm{tr}}\left[\hat{\rho}\ln[\hat{\rho}]\right]$ and together with Eq. (\ref{hamiltonian}) we obtain the variational equation for the density-matrix of the grand-canonical ensemble,
\begin{equation}
\delta\left\{{\rm{tr}}\left[\hat{S}\right]\,+\,\lambda_{1}{\rm{tr}}\left[\hat{H}\right]\,
+\,\lambda_{2}\,{\rm{tr}}\left[\hat{\rho}\,\hat{N}\right]\right\}\,=\,0\quad.
\label{variationaleq}\end{equation}
The Lagrange-parameters $\lambda_{i}$ in Eq. (\ref{variationaleq}) ensure that the system is constrained with respect to the conservation of the energy and the particle number. Variation of Eq. (\ref{variationaleq}) with respect to $\hat{\rho}$ yields
\begin{equation}
\ln[\hat{\rho}]\,+\,\lambda_{1}\left(\hat{H}_{0}\,+\,2\,g\,\hat{\rho}\right)\,+\,\lambda_{2}\,\hat{N}\,=\,0\quad.
\label{variationalresult}\end{equation}
Equation (\ref{variationalresult}) is solved in terms of the Lambert function $\Lambda$, which is related to the product-logarithm. In terms of the Lambert function $\Lambda$ the density-matrix of the interacting system is readily given by
\begin{equation}
\hat{\rho}\,=\,\Lambda\left(\hat{\rho}_{0}; \beta\,g\right)\,=\,\sum_{j=1}^{\infty}\frac{(-j)^{j-1}}{j!}(2\,\beta\,g)^{j-1}\,\hat{\rho}_{0}^{j}\quad.
\label{densitymatrix}\end{equation}
From the laws of thermodynamics we find the Lagrange-parameters with their familiar meaning $\lambda_{1}\,=\,\beta\,=\,1/k_{B}T$ and $\lambda_{2}\,=\,\beta\,\mu$, leading to the well-known density-matrix of the non-interacting system $\hat{\rho}_{0}\,=\,\exp[-\beta\,(\hat{H}_{0}\,-\,\mu\,\hat{N})]$. The symbolic equation (\ref{densitymatrix}) becomes meaningful when we calculate the partition function by applying the trace-operation
\begin{equation}
\mathcal{Z}\,=\,{\rm{tr}}\left[\hat{\rho}\right]\,=\,
\sum_{j=1}^{\infty}\frac{(-j)^{j-1}}{j!}(2\,\beta\,g)^{j-1}\,Z_{j}\quad.
\label{partitionfunction}\end{equation}
In Eq. (\ref{partitionfunction}) we have defined partial partitions
\begin{equation}
Z_{j}\,=\,{\rm{tr}}\left[\hat{\rho}^{j}\right]\,=\,\prod_{p}\,\frac{1}{1\,-\,e^{-\,j\,\beta(\epsilon_{p}\,-\,\mu)}}\quad.
\label{partialpartition}\end{equation}
The partial partitions in Eq. (\ref{partialpartition}) describe the degree of the interaction of the system according to the power-law expansion of the Lambert function Eq. (\ref{densitymatrix}). Quite surprisingly, the partial partitions have the same structure as in the non-interacting case, revealing the presence of particle-interaction only through the additional factor $j$. This special property is a direct consequence of the contact interaction assumed to be valid in the BEC case. When interactions are  non-local the density-matrix is also non-local, inducing tremendous complications to Eq. (\ref{densitymatrix}). The special property of Eq. (\ref{partialpartition}) can be understood by recalling the general property of the density-matrix, namely
\begin{equation}
{\rm{tr}}[\hat{\rho}]\,\geq\,{\rm{tr}}[\hat{\rho}^{2}],\quad {\rm{tr}}[\hat{\rho}]^{2}\,\geq\,{\rm{tr}}[\hat{\rho}^{2}]\quad.
\label{cauchyschwarz}\end{equation}
The inequalities Eq. (\ref{cauchyschwarz}) follow directly from the Cauchy-Schwarz inequality and can be extended to general powers. Thus, the partial partitions Eq. (\ref{partialpartition}) obey the relation $Z_{j}\,\geq\,Z_{j+1}$.

We rewrite the partial partitions $Z_{j}$ by introducing partial grand-canonical potentials $\Phi_{j}$
\begin{eqnarray}
Z_{j}\,&=&\,\exp\left[-\beta\,\Phi_{j}\right],\nonumber\\
\Phi_{j}\,&=&\,
\beta^{-1}\,\sum_{p}\,\ln\left[1\,-\,e^{-j\,\beta\,(\epsilon_{p}\,-\,\mu)}\right],
\label{partialgcp}\end{eqnarray}
which will prove useful for the following calculations. We proceed by calculating the grand-canonical potential $\Xi\,=\,-\,\beta^{-1}\,\ln[\mathcal{Z}]$, reading
\begin{eqnarray}
\Xi\,&=&\,\Phi_{1}\,-\,\beta^{-1}\,\sum_{k=1}^{\infty}\frac{(-1)^{k+1}}{k}\,\Xi_{k},\quad{\rm{with}}\nonumber\\
\Xi_{k}\,&=&\,\left(\sum_{j=2}^{\infty}\frac{(-j)^{j-1}}{j!}(2\,\beta\,g)^{j-1}\exp\left[\beta\,\Delta\Phi_{1,j}\right]\right)^{k}, 
\label{gcp}\end{eqnarray}
where we have separated the non-interacting part giving the first term and have already expanded the remaining logarithm into a power-series with the convenient notation $\Delta\Phi_{1,j}\,=\,\Phi_{1}\,-\Phi_{j}$. 

So far, our calculations have been carried out on a level of full generality. This allows us to apply our above theory to any bosonic quantum system where the free spectrum $\epsilon_{p}$ or, equivalently, the free density of states $D(\epsilon)$ is known.

A system of considerable interest in the context of BECs is the harmonic oscillator. In three dimensions, its density of states is given by \cite{Bag},
\begin{equation}
D(\epsilon)\,=\,\frac{\epsilon^{2}}{2\,\hbar^{3}\,\bar{\omega}^{3}}\quad,
\label{densityofstates}\end{equation}
where in Eq. (\ref{densityofstates}) $\bar{\omega}$ is the geometric mean of all frequencies $\bar{\omega}\,=\,(\omega_{1}\,\omega_{2}\,\omega_{3})^{1/3}$. When Eq. (\ref{densityofstates}) is inserted in Eq. (\ref{partialgcp}) and the sum in Eq. (\ref{partialgcp}) is taken in its integral limit which applies for large particle numbers, the partial grand-canonical potential is easily calculated, giving
\begin{equation}
\beta\,\Phi_{j}\,=\,-\frac{(k_{B}\,T)^{3}}{(\hbar\,\bar{\omega})^{3}}\frac{{\rm{Li}_{4}}\left[e^{j\,\beta\,\mu}\right]}{j^{3}}\quad,
\label{gcpho}\end{equation}
where the symbol ${\rm{Li}_{n}}$ in Eq. (\ref{gcpho}) is the usual notation for the poly-logarithmic function of order $n$. Knowing the partial partitions explicitly, the occupation number of the excited states can be obtained by the laws of thermodynamics
\begin{eqnarray}
N_{{\rm{ex}}}\,&=&\,-\,\left(\frac{\partial\,\Xi}{\partial\,\mu}\right)_{T}\,\nonumber\\
&=&\,\frac{(k_{B}\,T)^{3}}{(\hbar\,\bar{\omega})^{3}}\,\zeta(3)\,\chi\left(\frac{(k_{B}\,T)^{3}}{(\hbar\,\bar{\omega})^{3}}; \frac{g}{k_{B}\,T}\right)\quad.
\label{number}\end{eqnarray}
In Eq. (\ref{number}) $\zeta(3)$ is the Riemann zeta function, and the function $\chi$ describes the effect of the particle-interaction on the occupation of the excited states. 
\begin{figure}[t]\centering\hspace{-1.21cm}
\rotatebox{-90.0}{\scalebox{0.35}{\includegraphics{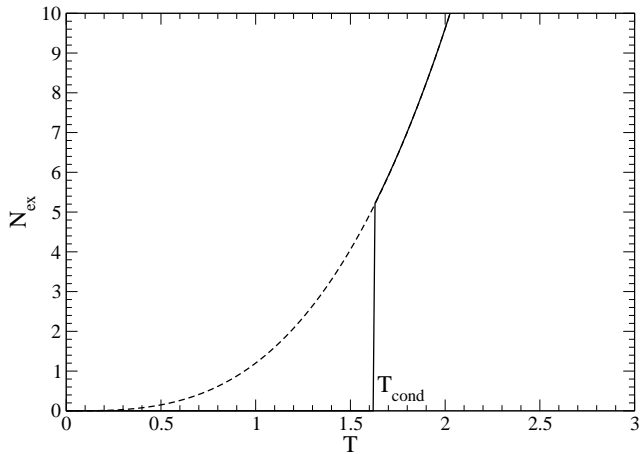}}}
\caption{\footnotesize{Occupation numbers $N_{{\rm{ex}}}$ for all excited states for the interacting case (full line) and the non-interacting case (dashed line). To illustrate the influence of the particle-interaction, we set $k_{{\rm{B}}}\,=\,\hbar\,=\,\bar{\omega}\,=\,1$ and choose $g\,=\,0.3$. The temperature $T_{{\rm{cond}}}$ marks the point where the BEC phase-transition occurs.}}
\label{OccupationNumber}\end{figure}
In Fig. 1 we have plotted the behaviour of Eq. (\ref{number}) and its non-interacting counterpart in arbitrary units for comparison. Clearly visible is a sudden fall-off of the number of occupied excited states at a certain temperature, that we will call the condensation temperature $T_{{\rm{cond}}}$ from now on. When the system has reached this temperature from above, all excited states experience a sudden deoccupation, such that the only occupied state left is indeed the ground-state of the system. This special behaviour of Eq. (\ref{number}) reveals the Bose-Einstein phase-transition for interacting systems not to be smooth. Our result is in sharp contrast to the Bose-Einstein phase-transition predicted for non-interacting systems, where the occupation number of excited states only vanishes as the temperature $T$ reaches zero, e.g. \cite{deG, Bag}. Strictly speaking, such behaviour does not contain any kind of phase-transition.

From the behaviour of Eq. (\ref{number}) in Fig. 1 it becomes clear that, when $T_{{\rm{cond}}}$ is known, the occupation number is sufficiently described by a step-function
\begin{equation}
N_{{\rm{ex}}}\,=\,\frac{(k_{B}\,T)^{3}}{(\hbar\,\bar{\omega})^{3}}\,\zeta(3)\,\theta\left[T\,-T_{{\rm{cond}}}\right]\quad. 
\label{numberstepfunction}\end{equation}
It is clear from Eq. (\ref{numberstepfunction}) that the existence of the condensation temperature $T_{{\rm{cond}}}$ does not harm the meaning of the critical temperature $T_{{\rm{c}}}$, see e.g. \cite{Bag}, as it should. The critical temperature remains to be the temperature at which, when approached from above, the ground-state of the system starts to get occupied. As a matter of fact, this occupation of the ground-state does not depend on the presence of particle-interaction and is a pure effect of cooling down a bosonic system. This explains the success of the non-interacting grand-canonical ensemble in this point. Only when $T_{{\rm{cond}}}$ is reached the particle-interaction gives rise to a sudden completion of the ground-state occupation. Thus, one could speak of a sliding phase-transition in the BEC-case, starting at the critical temperature $T_{{\rm{c}}}$ and finishing at the condensation temperature $T_{{\rm{cond}}}$.

According to Eq. (\ref{numberstepfunction}) the law for the condensed particle fraction stays in a form similar to the known one,
\begin{equation}
\frac{N_{0}}{N}\,=\,1\,-\,\frac{T^{3}}{T_{{\rm{c}}}^{3}}\,\theta[T\,-T_{{\rm{cond}}}]\quad. 
\label{condensedfraction}\end{equation}
\begin{figure}[t]\centering\hspace{-1.21cm}
\rotatebox{-90.0}{\scalebox{0.35}{\includegraphics{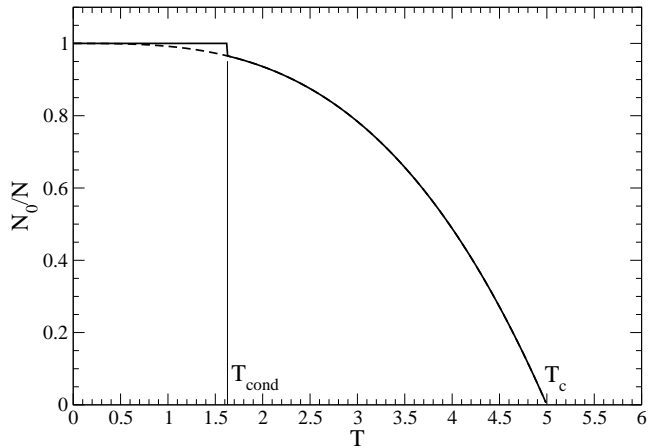}}}
\caption{\footnotesize{Condensed fraction of particles for the interacting case (full line) and the non-interacting case (dashed line) in arbitrary units for $g\,=\,0.3$ and a critical temperature $T_{{\rm{c}}}\,=\,5$. The vertical line marks the condensation temperature $T_{{\rm{cond}}}$.}}
\label{CondensedFraction}\end{figure}
The influence of $T_{{\rm{cond}}}$ can be seen from Fig. 2, where we have plotted the condensed particle fraction illustratively in arbitrary units. For temperatures $T_{{\rm{cond}}}\,<\,T\,\leq\,T_{{\rm{c}}}$ the behaviour of Eq. (\ref{condensedfraction}) shows the well-known power-law tail, e.g. \cite{Bag}, but for temperatures $0\,\leq\,T\,\leq\,T_{{\rm{cond}}}$, according to Eqs. (\ref{number}, \ref{numberstepfunction}) the number of excited particles falls off to zero and every particle is to be found in the ground state.

We now discuss our results in their relation to experiments that realize BECs. The influence of the particle-interaction in Eq. (\ref{number}) is described by the ratio $g/(k_{B}\,T)$. When we scale the system in units of its critical temperature $T_{{\rm{c}}}$ by introducing $T\,=\,\tau\,T_{{\rm{c}}}$ we can define an effective coupling constant
\begin{equation}
g_{{\rm{eff}}}\,=\,\frac{g}{k_{B}\,T_{{\rm{c}}}}\quad.
\label{effectivecoupling}\end{equation}
When Eq. (\ref{effectivecoupling}) is inserted into Eq. (\ref{number}), we can cast it into
\begin{equation}
N_{{\rm{ex}}}\,=\,\tau^{3}\,r^{3}\,\zeta(3)\,
\chi\left(\tau^{3}\,r^{3}; \frac{g_{{\rm{eff}}}}{\tau}\right),\quad r\,=\,\frac{k_{B}\,T_{{\rm{c}}}}{\hbar\,\bar{\omega}}\quad.
\label{numbertau}\end{equation}
With the value of $g_{{\rm{eff}}}$ from prominent experiments, e.g. \cite{Anderson, Davies, Ensher, Bradley, Sackett}, we find the interesting fact that all experiments have in common an extremely small value of $g_{{\rm{eff}}}$. For a quantitative illustration we take the JILA experiment \cite{Ensher}, where ${}^{87} {{\rm{Rb}}}$ atoms where Bose-Einstein condensed. The critical temperature in \cite{Ensher} was determined to be $T_{{\rm{c}}}\,=\,280\,{\rm{nk}}$ and the s-wave scattering length is found in \cite{Jul}, $a_{{\rm{int}}}\,=\,(103\,\pm\,5)\,\times\,a_{\rm{Bohr}}$. From this data follows a critical thermal energy $k_{{\rm{B}}}\,T_{{\rm{c}}}\,=\,3.86\,\times\,10^{-30}\,{\rm{J}}$ and an interaction energy of $g\,=\,5.27\,\times\,10^{-51}\,{\rm{J}}$. The effective coupling thus gives $g_{{\rm{eff}}}\,=\,1.36\,\times\,10^{-21}$, which is extremely small. Only when the dimensionless temperature $\tau$ reaches a magnitude of $\tau\,\approx\,10^{-19}\,-\,10^{-21}$, the effect of particle-interaction on the occupation number becomes visible. Thus, in up to date experimental setups, the thermal energy at the critical temperature $T_{{\rm{c}}}$ exceeds the interaction-energy $g$ by orders of magnitude. This makes the condensation temperature $T_{{\rm{cond}}}$ to lie only slightly above the absolute zero-point and the gas behaves effectively as a free gas with negligible influence of its internal interaction as long as $T_{{\rm{cond}}}$ close to the absolute zero-point has not been reached. This remarkable fact is the reason why, besides tiny corrections due to finite size effects, e.g. \cite{Gross}, no serious deviations - clearly distinct from systematic errors due to the measurement technique - from the predictions of the free grand-canonical ensemble have been detected in experiments so far. From technical remarks on the experimental feasibility, e.g. \cite{Ensher}, concerning especially measurements at very low temperatures close to the absolute zero-point we conclude, that the strongly increasing signal-to-noise ratio in this regime is responsible for the experimental inaccessibility of the extreme low-temperature region. However, a measurement of the condensed fraction of particles in a system that is close to a Feshbach-resonance, increasing $g$ by orders of magnitude, should make the BEC phase-transition become visible as predicted by Eqs. (\ref{number}, \ref{condensedfraction}).

\section{Influence of the environment}
Above we have mentioned experimental limitations for the access of the regime of extreme low temperatures. As no macroscopical system can be regarded as isolated, we proceed now by including effects, that may arise by the coupling of the condensate to a thermal environment, where at least black-body radiation is present. It is common sense in the BEC-field, and confirmed by experiment so far, that black-body radiation does not affect the condensation process, e.g. \cite{Cornell}. However, this changes when effects of non-linear ensembles are to be studied, as we will show now. Our ansatz follows the spirit of typical system-bath models, e.g. \cite{CalLeg, Lop, Ull}, but will be adopted to the requirements of our system. The simplest possible interaction between the system (1) and the bath (2) is given by a Hamiltonian
\begin{equation}
\hat{H}_{{\rm{int}}}\,=\,\kappa\,\hat{\rho}_{1}\,\hat{\rho}_{2}\quad.
\label{interactionhamiltonian}\end{equation}
The coupling-constant $\kappa$ has the meaning of some transition-rate. The full Hamiltonian is given by
\begin{equation}
\hat{H}\,=\,\left(\hat{H}_{0}\,+\,g\,\hat{\rho}_{1}\right)_{1}\,+\,\hat{H}_{2}\,+\,\kappa\,\hat{\rho}_{1}\,\hat{\rho}_{2}\quad.
\label{hamiltonian1}\end{equation}
When we discard the back-reaction from the system on the environment, we obtain the effective Hamiltonian by tracing out all contributions of the bath, giving
\begin{eqnarray}
\hat{H}_{1}^{\rm{eff}}\,&=&\,{\rm{tr}}[\hat{H}\,\hat{\rho}_{2}]\Rightarrow\nonumber\\
&=&\,\hat{H}_{0}\,+\,E_{2}\,+\,\left(g\,+\,\frac{\mathcal{Z}_{2}^{(2)}}{\left(\mathcal{Z}_{2}^{(1)}\right)^{2}}\,\kappa\right)\,\hat{\rho}_{1}\quad.
\label{effectivehamiltonian}\end{eqnarray}
The thermal energy and the partition-functions of the black-body radiation can be found in any textbook on statistics, e.g. \cite{LanLif}. We have $E_{2}(T_{2})\,=\,4\,\sigma/c\,T_{2}^{4}$, where $\sigma$ is the Stefan-Boltzmann constant and $c$ is the speed of light. Further, from our calculation above, Eqs. (\ref{partialpartition}, \ref{gcpho}), we can get for the partitions
\begin{eqnarray}
\mathcal{Z}_{2}^{(1)}\,=\,\exp\left[\frac{4\,\sigma}{3\,c\,k_{\rm{B}}}\,T_{2}^{3}\right],\quad\nonumber\\ 
\mathcal{Z}_{2}^{(2)}\,=\,\exp\left[\frac{\sigma}{6\,c\,k_{\rm{B}}}\,T_{2}^{3}\right]\quad.
\label{partitionsbath}\end{eqnarray}
From the partitions Eq. (\ref{partitionsbath}) it can be seen that for $T_{2}\,\rightarrow\,0$ the influence of the environment on the coupling vanishes. Finally, we introduce an effective coupling
\begin{equation}
\Gamma(g, T_{2}, \kappa)\,=\,g\,+\,R(T_{2})\,\kappa\quad,
\label{couplingenvironment}\end{equation}
where we have defined the ratio
\begin{equation}
R(T_{2})\,=\,\frac{\mathcal{Z}_{2}^{(2)}}{\left(\mathcal{Z}_{2}^{(1)}\right)^{2}}\,=\,\exp\left[-\frac{5}{2}\frac{\sigma}{c\,k_{\rm{B}}}\,T_{2}^{3}\right]\quad.
\label{ratio}\end{equation}
From Eq. (\ref{ratio}) it can easily be read off that the coupling is only influenced by the environment for small, but not too small temperatures $T_{2}$. For higher temperatures Eq. (\ref{ratio}) declines fast and renders the system with the bare particle-interaction only.
With Eq. (\ref{couplingenvironment}) the density-matrix including environmental effects, according to Eq. (\ref{densitymatrix}), readily follows
\begin{equation}
\hat{\rho}_{1}\,=\,\sum_{j=1}^{\infty}\frac{(-j)^{j-1}}{j!}\left(2\,\frac{\Gamma}{k_{\rm{B}}T_{1}}\right)^{j-1}\,\hat{\rho}_{0}^{j}\quad,
\label{densitymatrixenvironment}\end{equation}
where the density-matrix of the free system now shows a thermal energy-shift
\begin{equation}
\hat{\rho}_{0}\,=\,\exp\left[-\beta_{1}\left(\hat{H}_{0}\,+\,E_{2}(T_{2})\,-\,\mu_{1}\hat{N_{1}}\right)\right]\quad.
\label{freedensitymatrixenvironment}\end{equation}
It is worth noting, that this thermal shift is present even without any explicit coupling between system and bath. According to Eq. (\ref{partitionfunction}) we find the partition-function
\begin{equation}
\mathcal{Z}_{1}\,=\,
\sum_{j=1}^{\infty}\frac{(-j)^{j-1}}{j!}(2\,\beta_{1}\,\Gamma)^{j-1}\,Z_{j}\,\exp[-j\,\beta_{1}\,E_{2}(T_{2})]\quad.
\label{partitionfunctionenvironment}\end{equation}
In our following discussion, we will focus on the influence of the thermal energy-shift, since the coupling has only a marginal effect. From Eq. (\ref{partitionfunctionenvironment}) it can be immediately read off that the thermal energy-shift, for high environmental temperatures, beats down the interacting part of the partition-function such that only the non-interacting part for $Z_{j\,=\,1}$ survives. The destruction of the non-linear behaviour is initiated when the exponent in Eq. (\ref{partitionfunctionenvironment}) becomes equal or larger than unity. Thus we obtain the condition
\begin{equation}
T_{2}\,=\,\left(\frac{c\,k_{\rm{B}}}{4\,\sigma}\,T_{1}\right)^{\frac{1}{4}}\quad.
\label{condition1}\end{equation}
When $T_{1}$ is taken as the temperature of the system $T_{1}\,\sim\,10^{-9}$ K, the highest allowed value for $T_{2}$ is of the order $10^{-5}$ K. From this rather strict condition follows, that the emergence of a non-linear ensemble does not only require a tuned bare interaction $g$, but also a cold environment. Tuning the bare interaction by a Feshbach-resonance without an appropriately prepared environment, one would gain nothing.
The occupation-number of the system above $T_{\rm{cond}}$ is only weakly influenced by the thermal energy-shift, making an interacting system to behave like a non-interacting system. The result of this is well-known: condensation is completed only when absolute zero temperature is reached. Thus, the thermal energy-shift destroys the existence of the condensation temperature $T_{\rm{cond}}$ and shifts it to zero.

The subtle influence of the black-body radiation on the structure of the ensemble is only one effect of a \emph{dirty} vacuum. Other dirty vacuum effects are discussed in \cite{Cornell} as miscellaneous small effects and all of them are under suspect to simply destroy the non-linear structure of the BEC-ensemble and let it appear as being the well-known linear grand-canonical ensemble.

\section{Conclusion}
We have used the maximum information principle (MIP) to include the contact-interaction present in a BEC-system into the statistical description of the grand-canonical ensemble in a non-perturbative way. The MIP provides a very general approach, because it requires only the most elementary knowledge of the properties of the system: the Hamiltonian, the condition of the particle-conservation and the symmetry of the many-particle wave-function. Our analysis of the occupation number revealed the existence of a Bose-Einstein phase-transition occurring at a condensation temperature $T_{{\rm{cond}}}$. At $T_{{\rm{cond}}}\,>\,0$ all excited levels become deoccupied and the occupation of the ground-state is completed. The existence of the condensation temperature $T_{{\rm{cond}}}$ does not affect the meaning of the well-known critical temperature $T_{{\rm{c}}}$, where, when approached from above, the occupation of the ground-state is initiated. The existence of both temperatures states a sliding character of the Bose-Einstein phase-transition, starting at $T_{{\rm{c}}}$ and being abruptly completed at $T_{{\rm{cond}}}\,>\,0$.

We have discussed that the experimental state-of-the-art is unable to access the extreme low-temperature regime where deviations from the free grand-canonical ensemble description become visible. The effective coupling constant $g_{{\rm{eff}}}$ gives two possible paths of how the extreme low-temperature regime may become accessible in experiments. The first path may be a further reduction of the critical temperature of the system, however, this goes necessarily along with a drastic reduction of the total number of particles, that can only be avoided by an extremely flat trapping potential. The second path may be a tuning of the system by a Feshbach-resonance to manipulate $g$ instead of $T_{{\rm{c}}}$. 

We also have discussed possible effects of the environment in the case of a dirty vacuum, that is poisoned by black-body radiation or other miscellaneous effects as discussed in \cite{Cornell}. Our discussion revealed, that tuning the particle-interaction by a Feshbach resonance might not be enough, in the experiment one has also to take care for an almost clean vacuum around the condensed atomic cloud. At least for blocking black-body radiation as good as possible, one could think of a second atomic system surrounding the system of interest in order to create a screening effect. When such an artificial environment could be cooled down along with the condensate, thereby following the above relation, Eq. (\ref{condition1}), a realistic chance would emerge for experimentally probing the non-linear structure of the BEC-ensemble above zero temperature.

Our result hopefully inspires future technical developments towards the experimental accessibility of the low-temperature regime, because the technical handling of a condensation temperature $T_{{\rm{cond}}}$ above the absolute zero-point would allow to study condensed atomic systems behaving effectively as being at the absolute zero-point although their true temperature is still finite. We close our discussion by the general remark that our analysis in the BEC-case could provide some similarities to the phenomenon of high-temperature superconductivity. Its realization, however, depends on future experimental developments.

\end{document}